\documentclass[reprint,amsmath,amssymb,aps,prb,]{revtex4-2}

\usepackage{graphicx}
\usepackage{amsmath}
\usepackage{xcolor}
\usepackage{dcolumn}
\usepackage{enumitem}
\usepackage{epstopdf}

\begin{document}

\title{Antiferromagnetic resonances in twinned EuFe$_2$As$_2$ single crystal.}

\author{I.~A.~Golovchanskiy$^{1,2,6}$, N.N.~Abramov$^2$, V.A.~Vlasenko$^3$, K.~Pervakov$^3$, I.V.~Shchetinin$^2$, P.S.~Dzhumaev$^4$, O.V.~Emelyanova$^{4,5}$, D.S.~Baranov$^{1,4,6,7}$, D.S.~Kalashnikov$^{1,6}$, K.B.~Polevoy$^{1,6}$, V.M.~Pudalov$^{3,8}$, V.S.~Stolyarov$^{1,2,6}$}

\affiliation{
$^1$ Moscow Institute of Physics and Technology, State University, 9 Institutskiy per., Dolgoprudny, Moscow Region, 141700, Russia; \\
$^2$ National University of Science and Technology MISIS, 4 Leninsky prosp., Moscow, 119049, Russia; \\
$^3$ Ginzburg Center for High Temperature Superconductivity and Quantum Materials, P.N. Lebedev Physical Institute of the RAS, 119991, Moscow, Russia; 
$^4$ National Research Nuclear University MEPhI (Moscow Engineering Physics Institute), 31 Kashirskoye Shosse, 115409 Moscow, Russia;
$^5$ Center for Energy Science and Technology, Skolkovo Institute of Science and Technology, Nobel str. 3, 121205 Moscow, Russia;
$^6$ Dukhov Research Institute of Automatics (VNIIA), 127055 Moscow, Russia; 
$^7$ Skobeltsyn Institute of Nuclear Physics, MSU, Moscow, 119991, Russia; 
$^8$ HSE University, 101000 Moscow, Russia.
}%

\begin{abstract}
In this work, we report ferromagnetic resonance spectroscopy of EuFe$_2$As$_2$ single crystals.
We observe ferromagnetic resonance responses, which are attributed to antiferromagnetic resonances of Eu sub-lattice with orthorhombic crystal structure and with different orientations of twin domains relative to the external field.
We confirm validity of the recently-proposed spin Hamiltonian with anisotropic Eu-Eu exchange interaction and biquadratic Eu-Fe exchange interaction. 
\end{abstract}

\maketitle

\section{Introduction}

The discovery of the iron pnictide high temperature superconductors has been one of the most striking discoveries in recent years in condensed-matter physics.
In this class of materials, intriguing properties were recently found in the subclass of compounds originating from the parent EuFe$_2$As$_2$ \cite{Ren_PRB_78_052501,Jeevan_PRB_78_052502,Jiang_NJP_11_025007,Xiao_PRB_80_174424,Xiao_PRB_81_220406,Zapf_PRL_113_227001,Maiwald_PRX_8_011011,Sanchez_PRB_104_104413}.
Superconductivity in EuFe$_2$As$_2$-based ferromagnetic superconductors emerges by doping it with phosphorus \cite{Ren_PRL_102_137002,Cao_JPCM_23_464204,Jeevan_PRB_83_054511,Nandi_PRB_89_014512,Stolyarov_SciAdv_4_eaat1061,Grebenchuk_PRB_102_144501}, or substituting europium layers with rubidium \cite{Liu_PRB_93_214503,Liu_PRB_96_224510,Smylie_PRB_98_104503,Kim_PRB,Kim_arXiv,Iida_PRB_100_014506,Ishida_PNAS_118_e2101101118}.
In these materials the Fe 3d orbitals exhibit a dual itinerant-localized magnetism, and,  simultaneously, participate in superconducting pairing.
When it comes to the interplay between the superconductivity and the ferromagnetism, the studies are mainly focused on their superconducting properties and on the physical origin behind the emergence of superconductivity.
In case of EuFeAs-based ferromagnetic superconductors the coexistence is considered between magnetic ordering of Eu$^{2+}$ ions with large spin number $S=7/2$ and superconducting ordering of Fe-3d electrons.

\begin{figure}[!ht]
\begin{center}
\includegraphics[width=0.65\columnwidth]{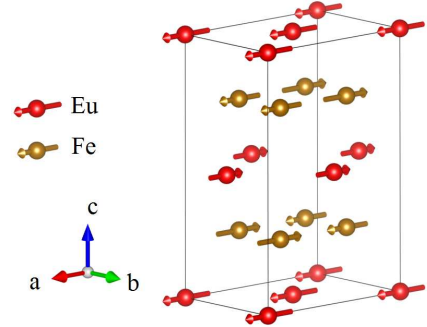}
\caption{
Magnetic crystal structure of EuFe$_2$As$_2$ (made using VESTA software).
The crystal structure of EuFe$_2$As$_2$ is orthorhombic with the space group $Fmmm$ \cite{Xiao_PRB_80_174424}.
Fe spin sub-lattice is in the spin density wave antiferromagnetic state aligned with the $a$ axis.
Eu spin sub-lattice is in the A-type antiferromagnetic state.
}
\label{coord}
\end{center}
\end{figure}

The EuFe$_2$As$_2$ compound itself is rich in various magnetic phenomena.
At about 190~K the crystal structure of EuFe$_2$As$_2$ undergoes the tetragonal-to-orthorhombic phase transition \cite{Xiao_PRB_80_174424,Maiwald_PRX_8_011011,Sanchez_PRB_104_104413} accompanied by the spin density wave antiferromagnetic transition in Fe sub-lattice.
The direction of spins in the spin density wave is locked along the longer $a$ crystal axis of the  orthorhombic structure (see Fig.~\ref{coord}).
At about 20~K the Eu subsystem undergoes the A-type antiferromagnetic transition \cite{Xiao_PRB_81_220406,Maiwald_PRX_8_011011,Sanchez_PRB_104_104413}.
The orthorhombic crystal structure is naturally subjected to twinning, but also to magnetostriction due to the movement of twin boundaries in response to changes of the magnetic field and magnetization of Eu sub-lattice \cite{Zapf_PRL_113_227001,Maiwald_PRX_8_011011,Sanchez_PRB_104_104413}. 
The latter indicates Eu-Fe exchange interaction.
This relocation of twin boundaries is one of the main obstructions for studies of magnetic ordering in EuFe$_2$As$_2$ with magnetization measurements: upon magnetization the variation in fraction of differently oriented twin domains additionally impacts the magnetization.
Only recently an adequate microscopic form of magnetic interactions in EuFe$_2$As$_2$ was established \cite{Maiwald_PRX_8_011011,Sanchez_PRB_104_104413}.
It includes the anisotropic Eu-Eu exchange interaction, bi-quadratic Eu-Fe exchange interaction, and implies the spin-flip transition in A-type antiferromagnetic Eu sub-lattice when the magnetic field is applied along the $a$ crystal axis. 
Metamagnetic transitions in EuFe$_2$As$_2$ \cite{Jiang_NJP_11_025007,Maiwald_PRX_8_011011,Sanchez_PRB_104_104413} are developed due to magnetization of twin domains with the $\pi/2$ difference in their mutual orientation.

In this work, we consider magnetization dynamics in single crystal EuFe$_2$As$_2$.
In general, ferromagnetic resonance (FMR) measurements are an ultimate tool for testing interlayer exchange interactions in various antiferromagnets \cite{Bogdanov_PRB_75_094425,Rezende_JAP_126_151101,Golovchanskiy_JAP}.
In addition, FMR studies are free from the twinning problem, since the ratio of differently oriented twin domains impact the intensity but not the position of resonance lines.  
By observing and analysing antiferromagnetic resonance spectral lines we have found that the FMR spectrum confirms the validity of the 3D-expanded version of the spin Hamiltonian proposed in Ref.~\cite{Sanchez_PRB_104_104413}.

\section{Experimental details}


EuFe$_2$As$_2$ single crystals were grown using the self-flux method, by analogy with previous works \cite{Pervakov_SUST_26_015008,Vlasenko_SUST_33_084009}. 
The initial high purity components of phase homogeneous EuAs (99.95\% Eu + 99.9999\% As) obtained by high-pressure technique, and preliminary synthesized precursor Fe$_2$As (99.98\% Fe+99.9999\% As) were mixed with a 1:6 molar ratio. 
The mixture in an alumina crucible was sealed in a niobium tube under residual argon pressure. The sealed container was loaded into a tube furnace with an argon atmosphere to prevent niobium oxidation. 
Then, the furnace was heated up to 1250$^\circ$C, held at this temperature for 24 h to homogenize melting, cooled down to 900$^\circ$C at a rate of 2$^\circ$C/h, and then cooled down to room temperature inside the furnace. 
Finally, crystals were collected from the crucible in an argon glove box.
Visually, as-grown bulk crystals demonstrate well-defined layered structure and their pliability for cleavage and exfoliation along the layering direction only.
XRD studies confirm alignment of $ab$ crystal planes within the layers, and orientation of c crystal axis across the layers.
Samples for ferromagnetic resonance spectroscopy were obtained by cleavage of as-synthesized bulk crystals.
Cleaved EuFe$_2$As$_2$ samples were of a few mm in size along $ab$ crystal planes, and about 50~$\mu$m thick along the $c$ crystal axis, which ensured the ``thin film'' geometry with defined crystal orientation.


Ferromagnetic resonance spectroscopy was performed using the VNA-FMR flip-chip approach\cite{Neudecker_JMMM_307_148,Kalarickal_JAP_99_093909}.
Cleaved EuFe$_2$As$_2$ sample was glued on top of the transmission line of coplanar waveguide.
The waveguide with impedance 50~Ohm and the width of the transmission line 0.5~mm is patterned on a Arlon AD1000 copper board and is equipped with SMP rf connectors.
The board with the sample is installed in a brass sample holder.
A thermometer and a heater are attached directly to the holder for precise temperature
control. 
The holder is placed in a home-made superconducting solenoid inside a closed-cycle cryostat (Oxford Instruments Triton, base temperature 1.2 K). 
Magnetic field is applied in-plane along the direction of the waveguide. 
The response of experimental samples is studied by analysing the transmitted microwave signal $S_{21}(f,H)$ with the VNA Rohde \& Schwarz ZVB20.
The setup allows to perform spectroscopy in the temperature range from 2~K up to 30~K, in the field range up to 1~T, and in the frequency range up to 26.5~GHz.

\section{Results and discussions}

\begin{figure*}[!ht]
\begin{center}
\includegraphics[width=0.99\columnwidth]{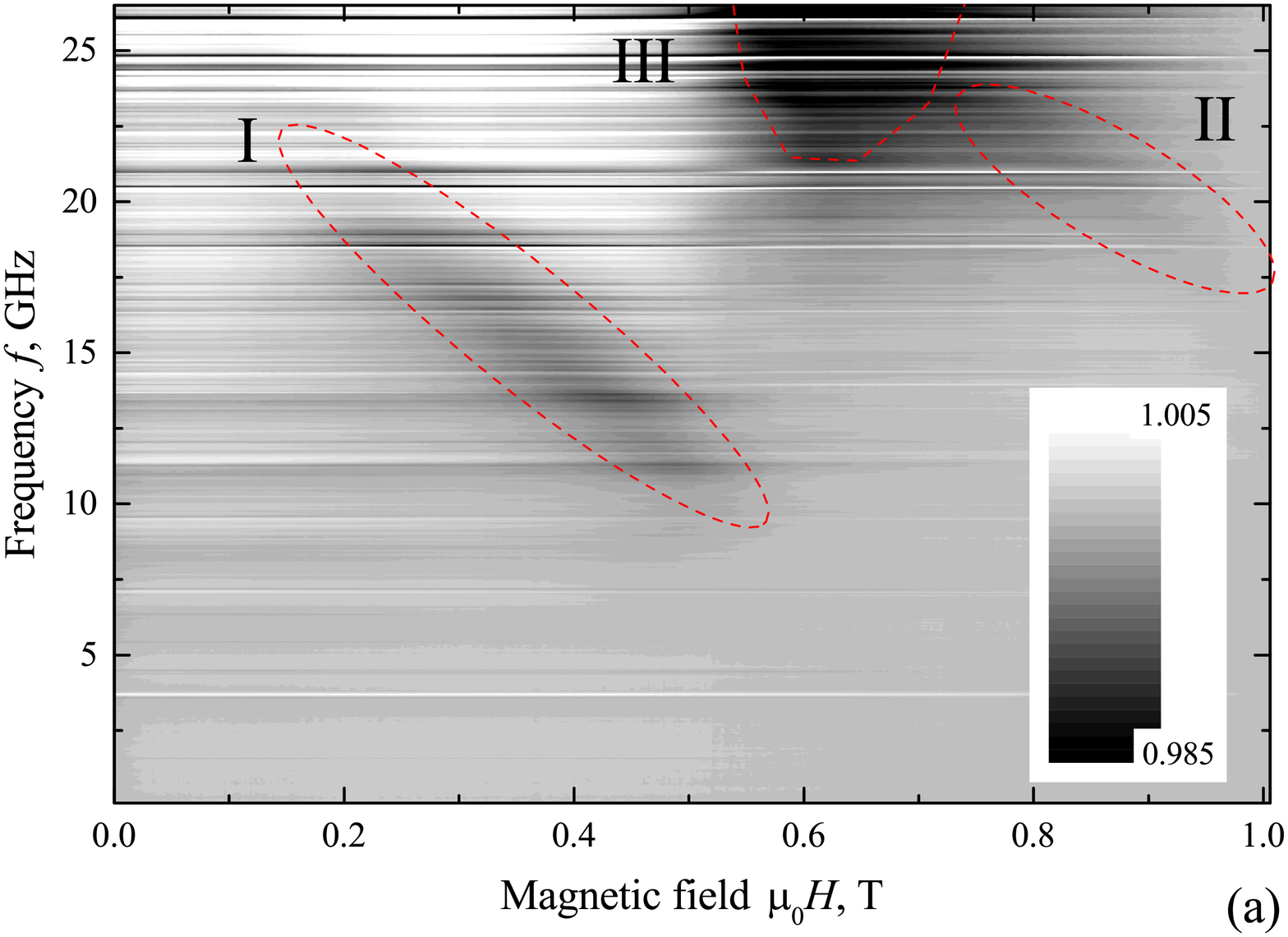}
\includegraphics[width=0.99\columnwidth]{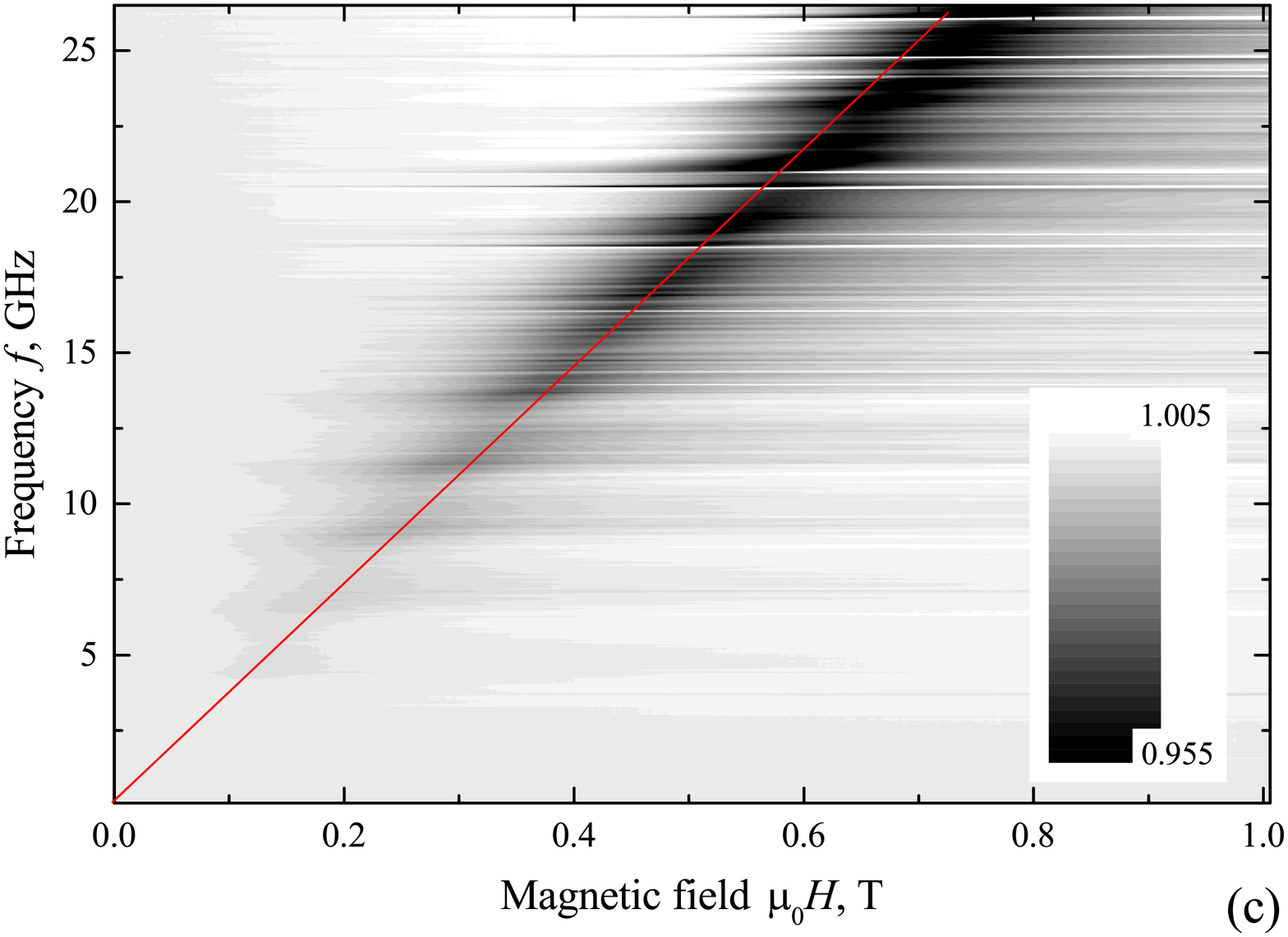}
\includegraphics[width=0.99\columnwidth]{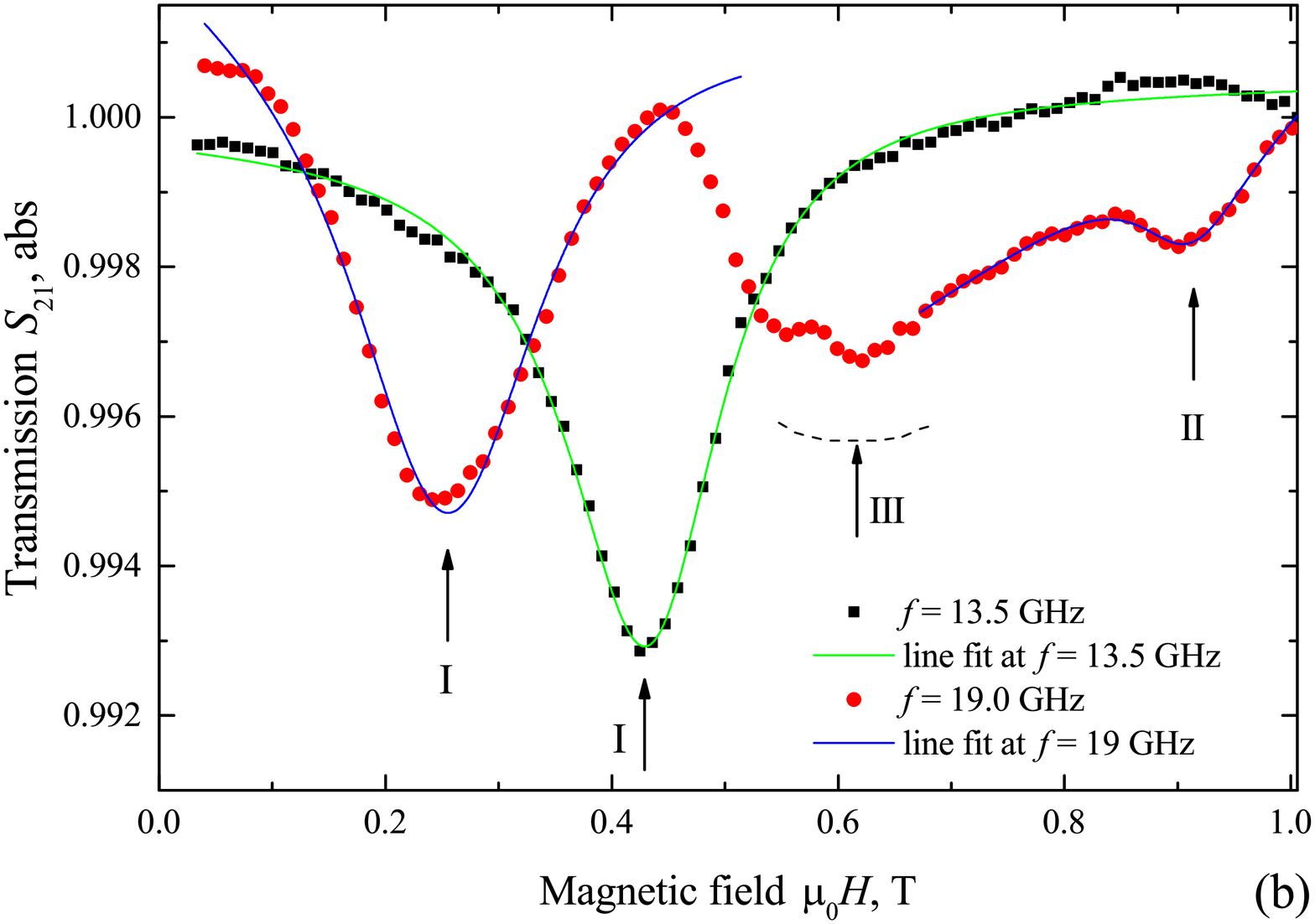}
\includegraphics[width=0.99\columnwidth]{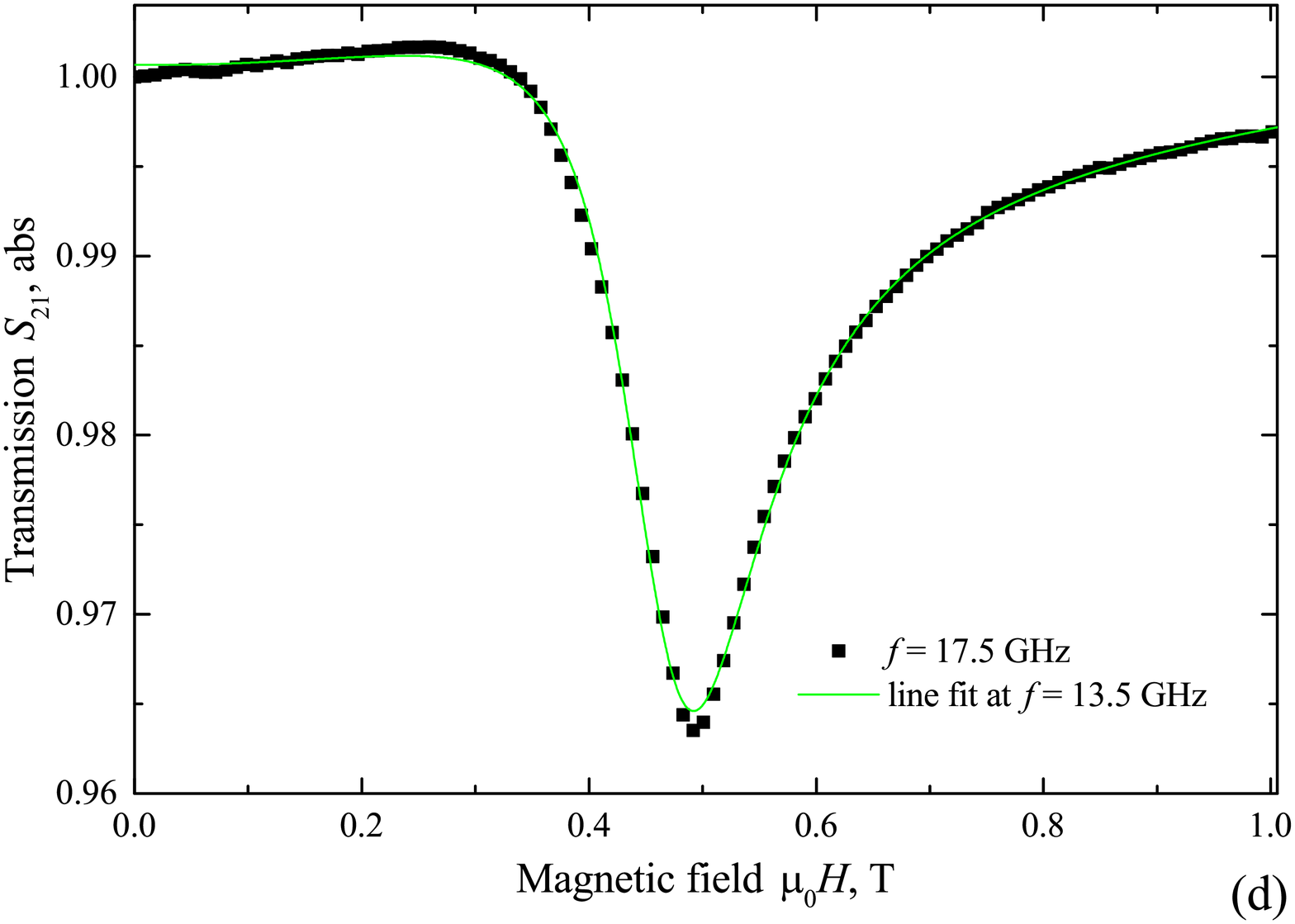}
\caption{a,c) FMR absorption spectra $S_{21}(f,H)$ of EuFe$_2$As$_2$ measured at $T=5$~K (a) and 20~K (c).
Magnetic field is applied in-plane along $ab$ crystal planes.
Circular red dashed lines in (a) highlight three absorption features.
Red line in (c) corresponds to the linear fit of the spectral line with the slope 36.6~GHz/T.
(b,d) Cross-sections $S_{21}(H)$ of spectra at $T=5$~K (b) and 20~K (d) at specified frequencies.
Solid lines in (b,d) show the fit of resonance lines, while arrows indicate corresponding absorption features.
}
\label{FMR_exp}
\end{center}
\end{figure*}

Figure~\ref{FMR_exp} shows FMR absorption spectra of EuFe$_2$As$_2$ sample measured at magnetic field applied in-plane along $ab$ layers at $T=5$~K and 20~K.
At 5~K (Fig.~\ref{FMR_exp}a) the spectrum contains three absorption features (encircled with red dashed lines):
(i) a resonance line at $0<\mu_0H\lesssim0.5$~T and $10<f<23$~GHz with a negative-slope linear field dependence, referred to as line I; 
(ii) a weaker resonance line at $\mu_0H\gtrsim0.7$~T and $18<f<22$~GHz, which also has approximately linear field dependence with a negative slope, referred to as line II; and
(iii) an absorption ``patch'' at $0.55\lesssim\mu_0H\lesssim0.7$~T and $f>20$~GHz, referred to as the absorption feature III.
A spectral line with the negative field-slope is commonly attributed to an antiferromagnetic resonance response.
Cross-sections $S_{21}(H)$ of the spectrum at 5~K and at selected frequencies are shown in Fig.~\ref{FMR_exp}b.
The $S_{21}(H)$ curves clearly indicate absorption peaks of resonance lines I and II.
Derivation of FMR frequency dependence on applied magnetic field $f_r(H)$ was performed by fitting $S_{21}(H)$ curves with the complex resonant susceptibility \cite{Kalarickal_JAP_99_093909}.

In contrast, at temperatures above 19~K the spectrum contains a single resonance line with approximately linear dependence of the resonance frequency on the magnetic field, corresponding to paramagnetic resonance of Eu spins.
Figure~\ref{FMR_exp}c shows the representative example of the spectrum at $T=20$~K.
The field slope of the paramagnetic resonance at 20~K $f_r/\mu_0H\approx36.6$~GHz/T is slightly higher than the gyromagnetic ratio of free electrons 28~GHz/T.
In addition, the resonance line shows some deviation from the linear behaviour. 
These effects are attributed to the residual susceptibility of the paramagnetic phase in vicinity to the Curie temperature \cite{MacNeill_PRL_123_047204} and should be understood as follows.
The magnetization of the paramagnetic phase is proportional to the applied field $M=\chi(T,H)H$. 
In the thin-film geometry the resonance frequency of the paramagnetic phase is still provided by the Kittel formula $2\pi f=\gamma\sqrt{H(H+M)}=\gamma H\sqrt{1+\chi(T,H)}$. 
At temperatures slightly above the Curie temperature, magnetization $\chi(T,H)H$ is comparable to the saturation magnetization at high fields and show nonlinear dependence on $H$. 
These factors result in larger slope of the linear fit and in some nonlinearity of the resonance line, which are observed in Fig.~\ref{FMR_exp}c. 
At higher temperatures $\chi(T,H)$ is gradually reduced and the resonance line approaches the conventional paramagnetic one, which is observed at 25 K and 30 K (see supplementary). 
Figure~\ref{FMR_exp}d shows the cross-section of the spectrum $S_{21}(H)$ at $f=17.5$~GHz, which contains a single resonance peak, and its fit with the complex susceptibility.


The spin configuration of Eu and Fe subsystems in EuFe$_2$As$_2$ and the twinning problem were studied extensively in a number of previous works with neutron scattering and XMCD measurements.  
As a consensus \cite{Sanchez_PRB_104_104413}, it is shown that at low temperature Fe sub-lattice is in the spin density wave antiferromagnetic state aligned with the longer side of orthorhombic lattice, while Eu sub-lattice has the A-type antiferromagnetic order (see Fig.~\ref{coord}).
Importantly, Eu-Fe exchange interaction results in anisotropic Eu-Eu exchange interaction and in development of the easy axis along the direction of the spin density wave (a-direction) due to bi-quadratic Eu-Fe exchange.
The total free energy of the spin configuration in a unit cell of Eu layers is \cite{Sanchez_PRB_104_104413}
\begin{equation}
\begin{aligned}
F={} & 2(J+W)e_{1x}e_{2x}+2(J-W)e_{1y}e_{2y}+2Je_{1z}e_{2z}+ \\
&  -8K_a\sum_{i=1}^{2} e_{ix}^2 + K_u\sum_{i=1}^{2} e_{iz}^2 - M_s\sum_{i=1}^{2} \vec{e}_i\vec{H},
\end{aligned}
\label{F_en}
\end{equation}
where $(e_{ix},e_{iy},e_{iz})$ is the unit vector of ferromagnetic moment of the Eu atomic layer in spherical coordinates $\vec{e}=\sin\theta\cos\phi\hat{x}+\sin\theta\sin\phi\hat{y}+\cos\theta\hat{z}$, 
$[\hat{x},\hat{y},\hat{z}]$ axes are aligned with $[a,b,c]$ crystal axes in Fig.~\ref{coord}, respectively,
the first three terms are the exchange interaction terms, which are anisotropic in $x$ and $y$ directions by the parameter $W$,
the forth term is the bi-quadratic Eu-Fe exchange interaction in a form of the $x$-axis uniaxial anisotropy,
the fifth term is the $z$-axis uniaxial anisotropy, 
and the last term is the Zeeman energy with the external field $\vec{H}$, which is applied in $ab$ plane at the angle $\phi_H$ with respect to the $a$-axis.
In comparison to Ref.~\cite{Sanchez_PRB_104_104413}, two terms $[2Je_{1z}e_{2z}]$ and $[K_u\sum_{i=1}^{2} e_{iz}^2]$ are added to complete the 3D representation of the free energy, while the Eu-Fe exchange interaction is redefined in the $x$-axis uniaxial form.

Following Ref.~\cite{Sanchez_PRB_104_104413}, exchange and anisotropy parameters of the free energy can be derived from saturation fields of the canted spin state and of the spin-flip transition as follows.
When magnetic field is applied along the $b$ crystal axis, magnetization of Eu occurs via spin canting (i.e., via the spin-flop phase) and the saturation field of the spin-flop phase is $H^{sat}_b=(4J+16K_a)/M_s$.
When magnetic field is applied along the $a$ crystal axis, magnetization saturation occurs via abrupt spin-flip transition and the saturation field (i.e., the spin-flip field) is $H^{sat}_a=2(J+W)/M_s$.
Notice that counter-intuitively $H^{sat}_a$ and $H^{sat}_b$ do not match each other even in the isotropic case of $W=0$, $K_a=0$.
This is the consequence of the spin-flip as the dominating magnetization process for the corresponding magnetic orientation \cite{Sanchez_PRB_104_104413}. 
The condition for the spin-flip transition is $J/(8K_a+W)<1$.
When magnetic field is applied at 45$^{\circ}$ with respect to $a$ or $b$ direction, the saturation field is $H^{sat}_{45}=4J/M_s$.
By expanding the treatment to the 3D case, the saturation field of the canted spin state for field orientation along the $c$ axis is $H^{sat}_c=(4J+2W+16K_a+2K_u)/M_s$.


The dependence of orientations of Eu magnetic moments on the magnetic field can be derived numerically by minimising the energy in Eq.~\ref{F_en} with predefined anisotropy and exchange parameters.
By knowing orientations of Eu magnetic moments, ferromagnetic resonance of Eu can be derived  using the Suhl-Smit-Beljers approach\cite{Suhl_PR_97_555,Smit_PRR_10_113} extended for the case of magnetization dynamics in coupled magnetic multilayers \cite{Zhang_PRB_50_6094,Schmool_JPCM_10_10679,Lindner_JPCM_15_S465,Golovchanskiy_JAP}.
With this approach the following set of equations of motion for magnetization vector in each Eu layer defines the collective dispersion of the spin system with orientation along $ab$ planes ($\theta_i=\pi/2$):
\begin{equation}
i\frac{\omega M_s}{\gamma}
\begin{bmatrix}
\delta\theta_1 \\
\delta\theta_2 \\
\delta\phi_1 \\
\delta\phi_2 \\
\end{bmatrix}
=
\begin{bmatrix}
0                     & 0                     & F_{\phi_1\phi_1} & F_{\phi_1\phi_2} \\
0                     & 0                     & F_{\phi_1\phi_2} & F_{\phi_2\phi_2} \\
-F_{\theta_1\theta_1} & -F_{\theta_1\theta_2} & 0                & 0                \\
-F_{\theta_1\theta_2} & -F_{\theta_2\theta_2} & 0                & 0                \\
\end{bmatrix}
\begin{bmatrix}
\delta\theta_1 \\
\delta\theta_2 \\
\delta\phi_1 \\
\delta\phi_2 \\
\end{bmatrix},
\label{Suhl}
\end{equation}
where $\delta\theta_i$ and $\delta\phi_i$ are components of small deviations of magnetization vector in spherical coordinates, $F_{\theta_i \theta_j}$ and $F_{\phi_i \phi_j}$ are corresponding second-order partial derivatives of the free energy (Eq.~\ref{F_en}), $\omega$ is the eigen-frequency of magnetization precession, and $\gamma=28$~GHz/T is the gyromagnetic ratio.
Diagonal terms $F_{\theta_i \phi_j}=0$ in Eq.~\ref{Suhl} due to the in-plane configuration of magnetization ($\theta_i=\pi/2$).
The expression $\sum\cos{(\phi_i-\phi_H)}\delta\phi_i$ corresponds to the dynamic susceptibility of the resonance.

\begin{figure}[!ht]
\begin{center}
\includegraphics[width=0.99\columnwidth]{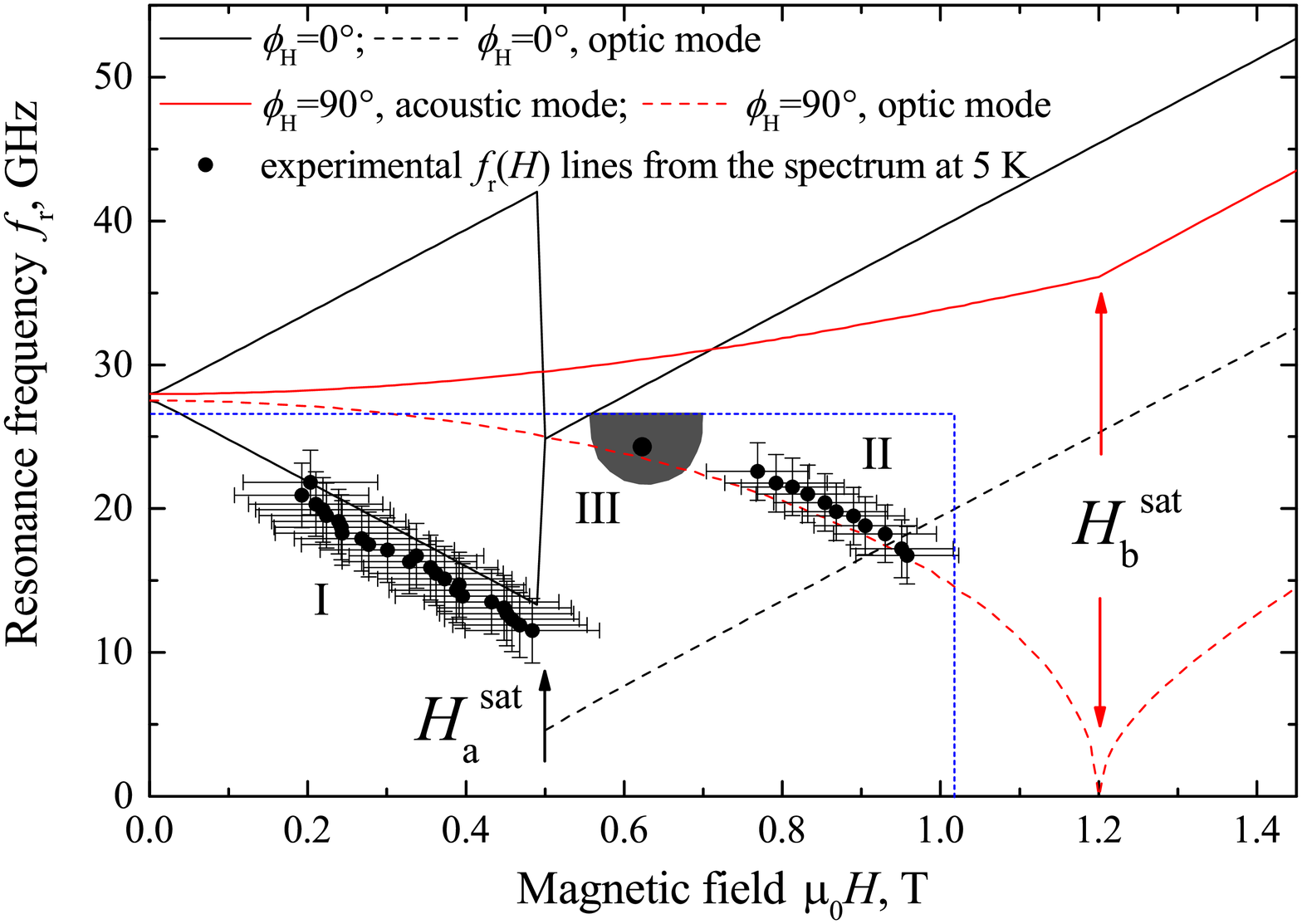}
\includegraphics[width=0.99\columnwidth]{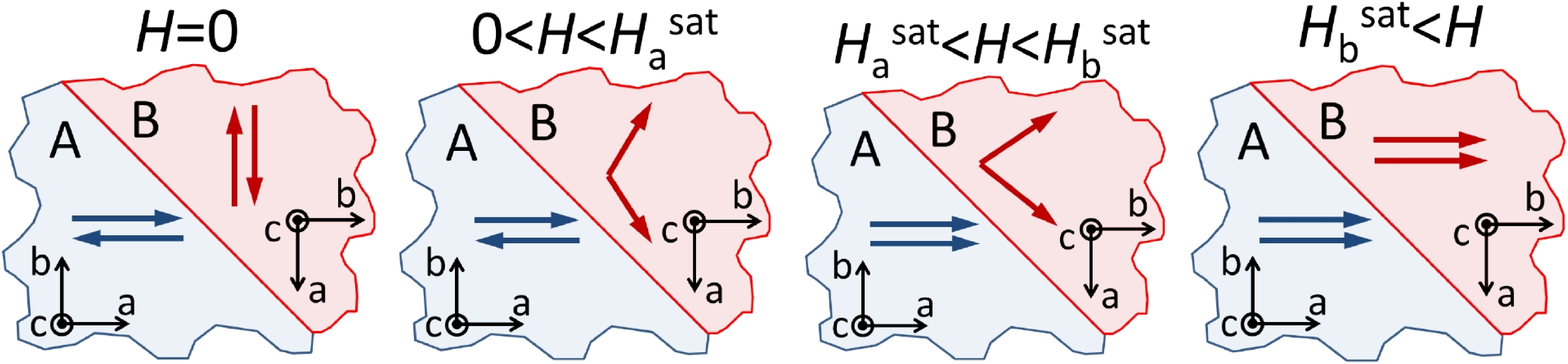}
\caption{
The plot shows experimental and theoretical dependencies of the ferromagnetic resonance frequency on the magnetic field $f_r(H)$.
Black lines show theoretical data for the A-domain with $\phi_H=0$.
Red lines show theoretical data for the B-domain with $\phi_H=\pi/2$.
Solid lines show either individual resonances of Eu sub-lattices, as in case of $\phi_H=0$ and $\mu_0 H<0.5$, or the collective acoustic response.
Dashed lines show the collective optical response. 
Arrows indicate transition fields $H^{sat}_a$ and $H^{sat}_b$.
Black squared dots show experimental $f_r(H)$ lines derived from the spectrum at 5 K.
Error bars indicate the line-width of experimental resonance lines $\Delta H\approx0.17$~T and $\Delta f\approx 4.5$~GHz for line I, and $\Delta H\approx0.13$~T and $\Delta f\approx 4$~GHz for line II.
Blue short-dashed line indicates the accessible range of the experimental setup.
Pictograms illustrate spin configurations in twinned domains at different fields.
A-domains (shown with blue) correspond to domains with the spin density wave axis ($a$-axis) aligned with the magnetic field.
B-domains (shown with red) correspond to domains with the spin density wave axis ($a$-axis) aligned perpendicular to the magnetic field.
}
\label{FMR_theory}
\end{center}
\end{figure}

Figure~\ref{FMR_theory} collects experimental and theoretical dependencies of FMR frequencies on magnetic field $f_r$(H).
In calculations we consider the in-plane magnetic field aligned with $ab$ crystal planes, with the angle $\phi_H$ relative to the $a$ crystal axis.
In accordance with the twinning domain concept \cite{Zapf_PRL_113_227001,Maiwald_PRX_8_011011,Sanchez_PRB_104_104413}, the sample also contains domains where the orientation of the magnetic field is $\pi/2-\phi_H$ relative to the $a$ crystal axis.

In general, our calculations showed that the value $\phi_H$ is close to 0, which indicates that the sample consists of domains whose $a$-axis is aligned with the magnetic field and domains whose $a$-axis is perpendicular to the magnetic field (A and B-domains, correspondingly, see illustrations in Fig.~\ref{FMR_theory}).
For A-domains at $H<H^{sat}_a$ the spectrum consists of two antiferromagnetic spectral lines with linear-in-field increasing (decreasing) resonant frequency, attributed to individual resonances of oppositely-aligned Eu spin sub-lattices.
At $H>H^{sat}_a$ the spin flip transition in A-domains occurs and the spectrum consists of two collective modes: the higher-frequency acoustic mode and the lower-frequency optical mode, both with linear field dependence.
Antiferromagnetic interaction between layers, which are magnetized to saturation, results in higher resonance frequency for the acoustic mode in comparison to the optical mode \cite{Lindner_JPCM_15_S465,Gurevich_book}.
For B-domains with the $b$-axis aligned with the magnetic field the spectrum also consists of two lines.
At $H<H^{sat}_b$ the spectrum of B-domains contains collective modes: higher-frequency acoustic mode with positive frequency dependence on magnetic field, and the lower-frequency optical mode with negative frequency dependence on magnetic field.
At $H>H^{sat}_b$ the spin-flip transition (saturation) occurs in B-domains manifested by a kink on both curves and both collective modes show positive dependence of frequency on magnetic field.

A rough numerical optimisation of magnetic parameters in Eqs.\ref{F_en} and \ref{Suhl} yields the following set of parameters, consistent with Ref.~\cite{Sanchez_PRB_104_104413}: $4J/M_s\approx0.8-0.9$~T; $2W/M_s\approx0.1-0.2$~T; $H^{sat}_a\approx0.45-0.55$~T; $H^{sat}_b=(4J+16K_a)/M_s\approx1.15-1.25$~T; $2K_u/M_s\approx0.2-0.3$~T; $|\phi_H|<5$~$^\circ$.
The large width of resonance lines and limited experiential range did not allow to perform more accurate optimisation of parameters.
Solid and dashed lines in Fig.~\ref{FMR_theory} show $f_r(H)$ obtained using Eqs.\ref{F_en} and \ref{Suhl} and the following set of parameters: $4J/M_s\approx0.8$~T; $2W/M_s\approx0.1$~T; $H^{sat}_a\approx0.5$~T; $H^{sat}_b=(4J+16K_a)/M_s\approx1.2$~T; $2K_u/M_s\approx0.25$~T; $|\phi_H|<5$~$^\circ$.
According to Fig.~\ref{FMR_theory}, the spectral line I corresponds to the resonance of Eu spins aligned against the external field in the domain having the angle $\phi_H=0$ with the external field.
The spectral line II corresponds to the optical antiferromagnetic response in the domain having the angle $\phi_H=\pi/2$ with the external field.
Its optical origin explains the low intensity in comparison with line I.

\begin{figure}[!ht]
\begin{center}
\includegraphics[width=0.99\columnwidth]{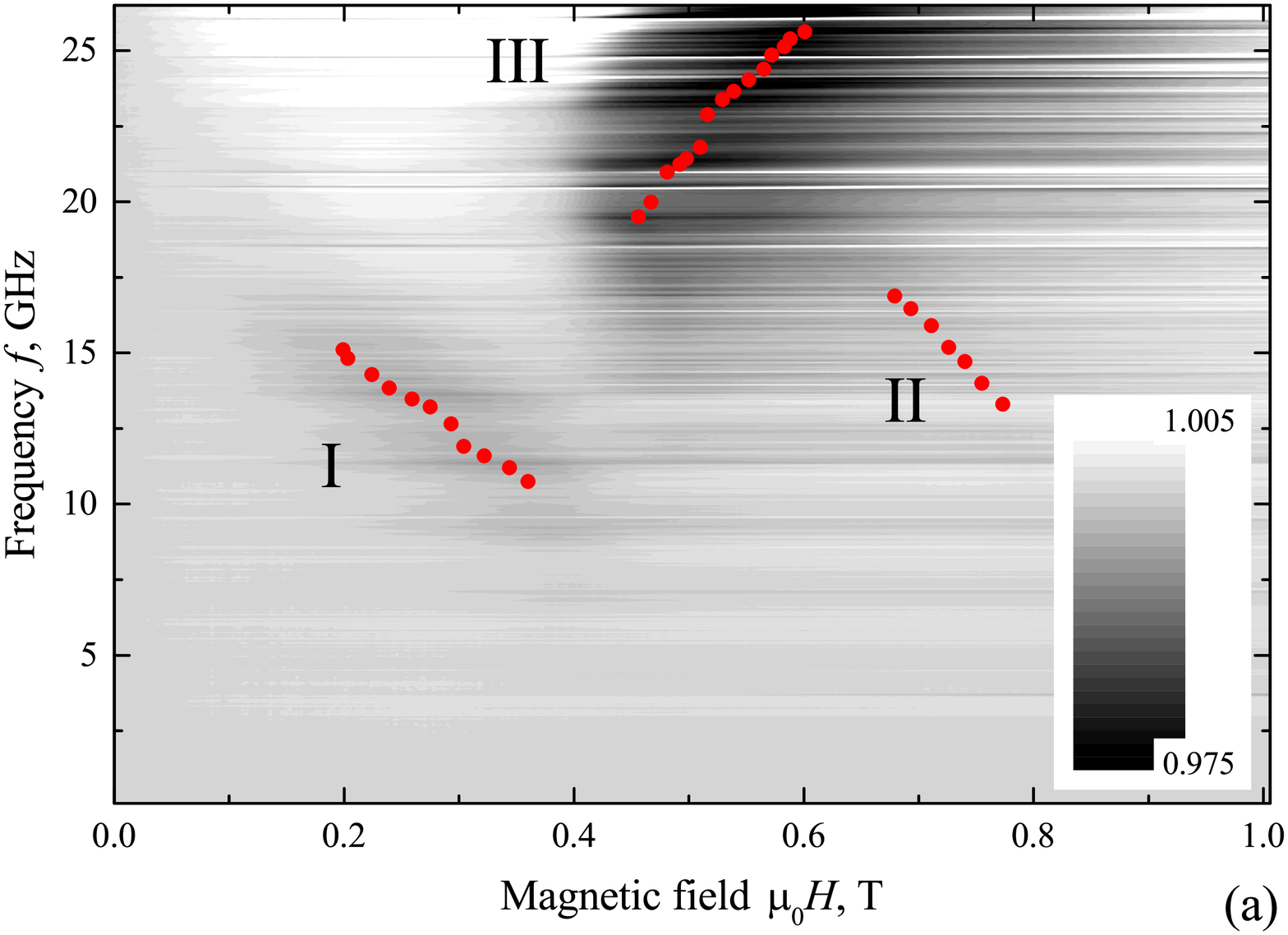}
\includegraphics[width=0.99\columnwidth]{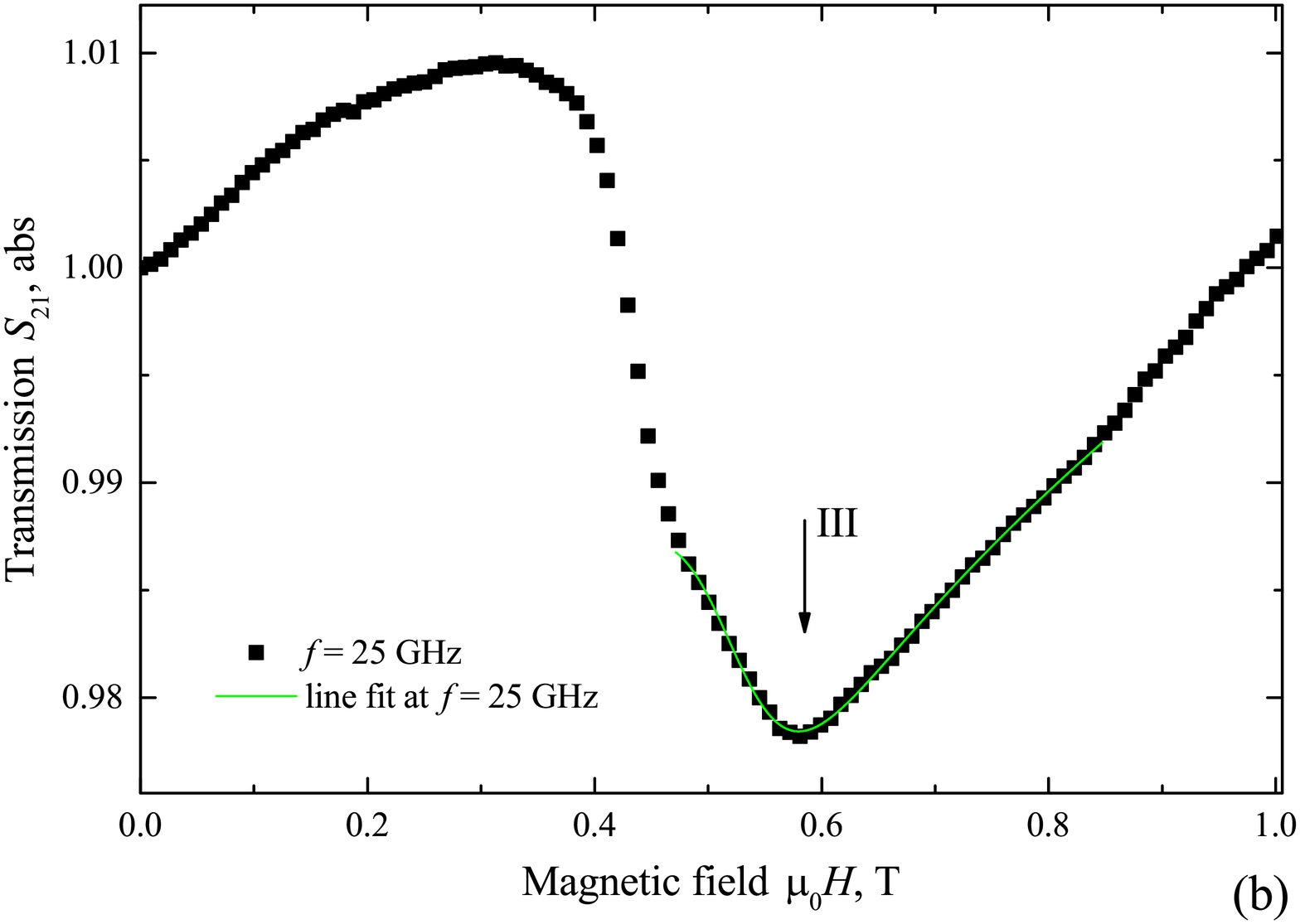}
\caption{a) FMR absorption spectra $S_{21}(f,H)$ of EuFe$_2$As$_2$ measured at $T=13$~K.
Magnetic field is applied in-plane along ab crystal planes.
Red dots show experimental resonance curves $f_r(H)$ obtained by fitting spectral lines.
b) Cross-section $S_{21}(H)$ of the spectrum at $T=13$~K at $f=25$~GHz.
Solid lines in (b) show the fit of the resonant feature.
}
\label{FMR_exp_13}
\end{center}
\end{figure}

The spectral feature III in Figs.~\ref{FMR_exp}a and ~\ref{FMR_theory} may be attributed to the optical antiferromagnetic response in the same B-domain.
In this case the change in the line intensity is the signature of the twin domain wall relocation when the fraction of these domains reduces with the magnetic field.
Alternatively, the spectral feature III may be a trace of the acoustic mode of the A-domain with $\phi_H=0$ and $H>H^{sat}_a$.
The origin of the spectral feature III can be established by studying temperature dependence of the spectrum.
Upon increasing the measurement temperature all spectral lines shift to lower frequencies, while the transition fields decrease (see supplementary).
Figure~\ref{FMR_exp_13}a shows FMR absorption spectrum of EuFe$_2$As$_2$ sample measured at magnetic field applied in-plane along $ab$ layers at $T=13$~K.
The spectrum contains the same three absorption features (experimental $f_r(H)$ dependencies are shown with red dots).
The resonance line I at $0.2<\mu_0H\lesssim0.4$~T and $10<f<15$~GHz corresponds to resonance of Eu spins aligned against the external field in the A-domain having the angle $\phi_H=0$ with the external field (see schematic images in Fig.~\ref{FMR_theory}).
A weak resonance line II at $0.7\lesssim\mu_0H\lesssim0.8$~T and $12<f<18$~GHz corresponds to the optical antiferromagnetic response in the B-domain having the angle $\phi_H=\pi/2$ with the external field.
The spectral feature III at $T=5$~K (Fig.~\ref{FMR_exp}a) is transformed at $T=13$~K into a clearly distinguishable resonance line at $0.45\lesssim\mu_0H\lesssim0.7$~T and $18<f<26$~GHz with positive-slope linear dependence $f_r(H)$, thus, manifesting the acoustic mode of A-domains with $\phi_H=0$ and $H>H^{sat}_a$.
Lines I and III indicate that the spin-flip field at $T=13$~K is reduced to $H^{sat}_a\approx0.4$~T as compared to $H^{sat}_a\approx0.5$~T at 5~K.
Figure~\ref{FMR_exp_13}b shows the cross-section of the spectrum $S_{21}(H)$ at $f=25$~GHz.
The cross-section indicates a drop of the transmission at $\mu_0H\approx0.4$~T.
This drop is attributed to the spin-flip transition of A-domains but is not related to a spin resonance process. 
At $\mu_0H>0.4$~T magnetization of A-domains is changed step-wise, which result in corresponding changes of the impedance of the transmission line and, consequently, in its transmission characteristics regardless a resonance process. 
In addition, the curve $S_{21}(H)$ shows a resonance peak III at $\mu_0H\approx0.58$~T.
The fit of resonance peaks with the complex susceptibility allowed to derive the resonance line $f_r(H)$, which is shown in Fig.~\ref{FMR_exp_13}a with red dots.

The overall correspondence between experimental and theoretical lines in Fig.~\ref{FMR_theory} and established nature of all lines in Figs.~\ref{FMR_theory} and \ref{FMR_exp_13} confirm the validity of the free energy relation in Eq.~\ref{F_en} with the anisotropic Eu-Eu exchange interaction and bi-quadratic Eu-Fe exchange interaction for EuFe$_2$As$_2$ compound together with the domain twinning concept of its orthorhombic crystal structure.

\begin{figure}[!ht]
\begin{center}
\includegraphics[width=0.99\columnwidth]{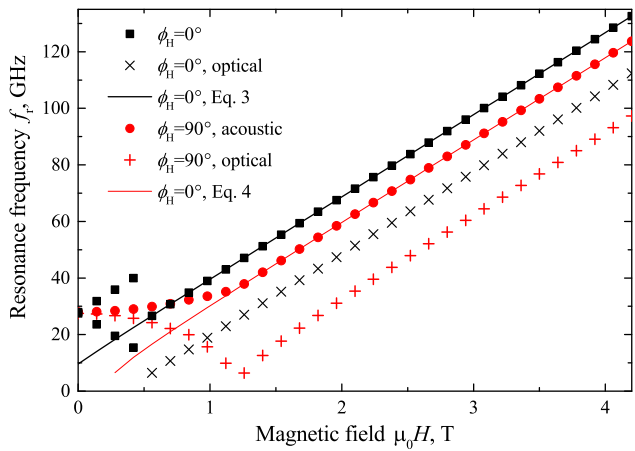}
\caption{Theoretical dependencies of the ferromagnetic resonance frequency on the magnetic field $f_r(H)$ obtained numerically with Eqs.~\ref{F_en},\ref{Suhl} using the same parameters as in Fig.~\ref{FMR_theory} (dots), and analytically with Eqs.~\ref{Kit_A},\ref{Kit_B} (solid lines).
}
\label{FMR_model}
\end{center}
\end{figure}

Once the validity of the Hamiltonian~\ref{F_en} and the origin of all three lines are established it is instructive to consider resonance properties of EuFe$_2$As$_2$ in more details.
First, it can be noticed that at zero field in Fig.~\ref{FMR_theory} resonance lines are split by $\Delta f_r\approx 0.5$~GHz.
In general, for two-sublattice antiferromagnets the zero-filed split occurs due to the two-axes anisotropy.
From Eq.~\ref{Suhl} at $H=0$ it can be derived that the split in FMR frequency is set by the difference between two eigen frequencies given by expressions
\begin{equation}
\begin{aligned}
\left(\frac{2\pi f_r}{\mu_0\gamma}\right)^2=(2W+16K_a+2K_u)(4J+16K_a), \\
\left(\frac{2\pi f_r}{\mu_0\gamma}\right)^2=(4J+2W+16K_a+2K_u)(4W+16K_a).
\end{aligned}
\label{AFMR}
\end{equation}
As follows from Eq.~\ref{AFMR}, in the case of EuFe$_2$As$_2$ the width of the split $\Delta f_r$ is also affected by the anisotropy in exchange interaction.
In the isotropic case, $W=0$ and $K_u=0$, both expressions relax to the known textbook split-less expression \cite{Rezende_JAP_126_151101,Gurevich_book}.
However, it should be noticed that $\Delta f_r$ is by far smaller that the linewidth of resonance lines, and, thus, can not be verified directly for EuFe$_2$As$_2$.

Next, we consider the resonance of magnetically-saturated A- and B-domains.
In terms of Eq.~\ref{F_en} by setting $\vec{e_1}=\vec{e_2}$ the Suhl-Smit-Beljers approach yields for a single magnetic layer
\begin{equation}
\left(\frac{2\pi f_r}{\mu_0\gamma}\right)^2=(H+16K_a-4W)(H+16K_a-2W+2K_u)
\label{Kit_A}
\end{equation}
for the higher-frequency acoustic mode of the A-domain at $H>H^{sat}_a$, and 
\begin{equation}
\left(\frac{2\pi f_r}{\mu_0\gamma}\right)^2=(H-16K_a+4W)(H+2W+2K_u)
\label{Kit_B}
\end{equation}
for the higher-frequency acoustic mode of the B-domain at $H>H^{sat}_b$. 
Figure~\ref{FMR_model} shows theoretical dependencies of the ferromagnetic resonance frequency on the magnetic field $f_r(H)$ obtained numerically with Eqs.~\ref{F_en},\ref{Suhl} using the same parameters as in Fig.~\ref{FMR_theory}, and analytically with Eqs.~\ref{Kit_A},\ref{Kit_B}.
The consistency between corresponding resonance lines confirms the validity of numerical studies.
Interestingly, the exchange anisotropy parameter $W$ enters both expressions, which can be used additionally in FMR determination of magnetic properties of complex antiferromagnets.

%
%

\section{Conclusion}

In conclusion, in this work we report ferromagnetic resonance spectroscopy of EuFe$_2$As$_2$ single crystals.
The spectrum reveals several resonant features attributed to antiferromagnetic resonances of Eu sub-lattice.
By employing the recently proposed spin Hamiltonian with anisotropic Eu-Eu exhange interaction and bi-quadratic Eu-Fe exchange interaction, the spectral features have been identified and attributed to antiferromagnetic and collective resonances of Eu layers in orthorhombic twinned crystal with different orientation of twin domains with respect to the external field.
The obtained magnetic parameters are quantitatively consistent with those reported previously, thus, confirming the complex biquadratic Hamiltonian for Eu spins in EuFe$_2$As$_2$ proposed earlier.

\section{Acknowledgments}

This work was supported by the Russian Science Foundation and by the Ministry of Science and Higher Education of the Russian Federation. 
Crystal synthesis was done using equipment from the LPI Shared Facility Center and was partially supported by the Russian Foundation for Basic Research.

\bibliographystyle{apsrev}
\bibliography{A_Bib_EuRbFeAs}

\end{document}